\documentclass{article}

\usepackage{arxiv}

\usepackage[utf8]{inputenc} 
\usepackage[T1]{fontenc}    
\usepackage{hyperref}       
\usepackage{url}            
\usepackage{booktabs}       
\usepackage{amsfonts}       
\usepackage{nicefrac}       
\usepackage{microtype}      
\usepackage{lipsum}

\usepackage{graphicx}
\usepackage{amsmath,amssymb}
\pdfoutput=1
\usepackage{changepage}
\usepackage{textcomp,marvosym}
\usepackage[right]{lineno}
\usepackage{microtype}
\DisableLigatures[f]{encoding = *, family = * }
\usepackage[table]{xcolor}
\usepackage{array}

\newcolumntype{+}{!{\vrule width 2pt}}

\newlength\savedwidth

\newcommand\thickhline{\noalign{\global\savedwidth\arrayrulewidth\global\arrayrulewidth 2pt}%
\hline
\noalign{\global\arrayrulewidth\savedwidth}}

\makeatletter

\title{Order flow in the financial markets from the perspective of the Fractional L\'evy stable motion}

\author{
  Vygintas~Gontis\thanks{Use footnote for providing further
    information about author (webpage, alternative
    address)---\emph{not} for acknowledging funding agencies.} \\
  Institute of Theoretical Physics and Astronomy\\
  Vilnius University\\
  Saul{\. e}tekio al. 3, 10257 Vilnius, Lithuania \\
  \texttt{vygintas@gontis.eu} \\
}

\begin{document}
\maketitle

\begin{abstract}
It is a challenging task to identify the best possible models
based on given empirical data of observed time series. Though the financial markets provide us with a vast amount of empirical data, the best model selection is still a big challenge for researchers. The widely used long-range memory and self-similarity estimators give varying values of the parameters as these estimators themselves are developed for the specific models of time series. Here we investigate from the general fractional L\'evy stable motion perspective the order disbalance time series constructed from the limit order book data of the financial markets. Our results suggest that previous findings of persistence in order flow could be related to the power-law distribution of order sizes and other deviations from the normal distribution. Still, orders have stable estimates of anti-correlation for the 18 randomly selected stocks when Absolute value and Higuchi's estimators are implemented. Though the burst duration analysis based on the first passage problem of time series and implemented in this research gives slightly higher estimates of the Hurst and memory parameters, it qualitatively supports the importance of the power-law distribution of order sizes.
\end{abstract}


\section{Introduction}
Power-law statistical properties are the characteristic feature of social systems. The financial markets are providing us with a vast amount of empirical limit order book (LOB) data that exhibit such power-law statistical properties as well \cite{Gould2013QF}. The long-range memory in finance and other natural and social systems is closely related to self-similarity and power-law statistical properties. The empirical data and observed statistical properties of volatility, trading activity, and order flow in the financial markets are still among the most mysterious features attracting the permanent attention of researchers \cite{Baillie1996JE,Engle2001QF,Plerou2001QF,Gabaix2003Nature,Ding2003Springer}. The widely used measures of long-range memory based on the self-similarity and power-law statistical properties are ambiguous as Markov processes can exhibit long-range memory properties, including slowly decaying auto-correlation \cite{Gontis2004PhysA,McCauley2006PhysA,McCauley2007PhysA,Micciche2009PRE,Micciche2013FNL,Ruseckas2011PRE,Kononovicius2015PhysA}. Various models, such as FIGARCH, FIEGARCH, LM-ARCH, and ARFIMA, including fractional noise, have been proposed for the volatility in the financial markets \cite{Ding1993JEmpFin,Baillie1996JE,Bollerslev1996Econometrics,Giraitis2009,Conrad2010,Arouri2012,Tayefi2012}.  

Our continuing research recently reviewed in \cite{Kazakevicius2021Entropy} raises the question of whether the observed long-range memory in social systems is a result of the actual long-range memory process or just a consequence of the non-linearity of Markov processes. Earlier, we have reduced the macroscopic dynamics of the financial markets to a set of stochastic differential equations (SDEs) able to reproduce empirical probability density function (PDF) and power spectral density (PSD) of absolute return \cite{Kononovicius2013EPL,Gontis2014PlosOne,Gontis2017PhysA,Gontis2018PhysA}. We used the same model to interpret the scaling behavior of volatility return intervals \cite{Gontis2016PhysA}. For the empirical analysis, it was helpful to employ the dependence of first passage time PDF on Hurst exponent $H$  \cite{Ding1995PhysRevE,Metzler2014Springer} in search of long-range memory properties for the volatility \cite{Gontis2017PhysA,Gontis2018PhysA} and for the order flow \cite{Gontis2020JStatMech} in the financial markets.
The proposed description is an alternative to the modeling incorporating fractional Brownian motion (FBM). Non-linear SDEs might be applicable in the modeling of other social systems, where models of opinion or populations dynamics lead to the macroscopic description by the non-linear SDEs \cite{Gontis2017Entropy}. Thus, the question of which model is best suited for interpreting empirical data is always a big challenge. The most general initial assumptions help search for the correct answer.

There is tremendous interest in long-range memory and self-similar processes, in particular FBM, fractional L\'evy stable motion (FLSM), and auto-regressive fractionally integrated moving average (ARFIMA) \cite{Burnecki2010PRE,Burnecki2014JStatMech,Burnecki2017ChaosSF}. These processes first of all serve for the modeling of systems with anomalous diffusion and expected fractional dynamics \cite{Klafter2012WorldScientific}. The first two models are self-similar with correlated increments.  The discrete-time ARFIMA process generalizes both models as in the limit it converges to either FBM or FLSM. Self-similar processes with non-Gaussian stable increments are essential for the modeling of social systems, where power-law distributions often interplay with auto-correlations \cite{Lillo2004SNDE,Bouchaud2004QF,Toth2015JEDC}. 

In the previous paper \cite{Gontis2020JStatMech} we investigated burst and inter-burst duration statistical properties of order disbalance time series seeking to confirm or reject the long-range memory in the order flow. Nevertheless, we have to admit that the necessary assumption of non-Gaussian increments was missing in that research. Here we analyze the same LOBSTER data of order flow in the financial markets \cite{Huang2011Lobster} from the perspective of FLSM and ARFIMA models seeking to identify the impact of increment distributions and correlations on estimated parameters of self-similarity.

The paper we organize as follows. In section \ref{FractionalTS} we introduce fractional time series and methods used to evaluate the scaling parameters. In section \ref{sec:data} we describe data sources of limit order books (LOB) and define order disbalance time series. In section \ref{sec:reults} we apply selected methods to the data and demonstrate the particularities of burst and inter-burst duration analysis. In the concluding part, we discuss the results and summarize the findings.

\section{Fractional time series}
\label{FractionalTS}
\subsection{Fractional Brownian motion}
\label{sec:FBM}
Fractional Brownian motion (FBM) was proposed as a generalization of the classical Brownian motion (BM) \cite{Mandelbrot1968SIAMR}. It serves as a model of the correlated time series with stationary Gaussian increments. FBM of index H (Hurst parameter) in the interval $0<H<1$  is
the mean-zero Gaussian process $B_H(t)$. 
The parameter H in FBM quantifies few essential properties: fractal behavior, long-range memory, and anomalous diffusion. We will argue later that this is not the case for the other more general modeling. Thus we assume that Hurst parameter $H$ is responsible for the fractal properties of the trajectories and simultaneously for its property of self-similarity. For any $a > 0$, $B_H (at)$ and $a^H B_H (t)$ have identical distributions. One can establish the relation with the fractal dimension of trajectories $D=2-H$ as well \cite{Magdziarz2013JPhysA}. In analogy to the notions used in fractal geometry, this type of process can be considered self-similar.

Very important property of many complex systems generating time series $X(t)$ is the anomalous diffusion characterized by the mean square displacement (MSD) growing as a power-law of time $\langle X^2(t)\rangle \sim t^{\lambda}$, where exponent $\lambda$ defines the memory parameter $d=(\lambda-1)/2$. For the FBM $d=H-1/2$ \cite{Klafter2012WorldScientific}. For $d<0$ one observes dynamics as sub-diffusion and for $d>0$ as super-diffusion. In experimental or empirical data analyses one usually deals with discrete-time sample data series ${X_i, i=0,1,...,N}$. We will use the sample MSD defined as
\begin{equation}
M_N(k)=\frac{1}{N-k+1} \sum_{i=o}^{N-k} (X_{i+k}-X_k)^2.
\label{eq:sample-MSD}
\end{equation}  
Note that MSD definition using ensemble average is valid for the FBM, while for the FLSM diverges \cite{Burnecki2010PRE}. The authors provide an evidence of FLSM non-ergodicity and that $M_N(k) \sim k^{\lambda}$, where $\lambda=2d+1$, for large $N$, $k$, and $N/k$. Thus the MSD sample analysis of time series with FLSM assumption becomes very important, estimating the memory parameter $d$.

We denote by ${Y_i=X_i-X_{i-1}, i=1,...,N}$ as increment process retrieved from the sample data series. In the case of FBM increment process is called fractional Gausssian noise (FGN). The long-range memory usually is defined through the divergence of autocovariance $\rho(k)$, \cite{Taqqu1995Fractals} 
\begin{equation}
\rho(k)=\frac{1}{N-k+1} \sum_{i=1}^{N-k+1} Y_i Y_{i+k}
\sim H(2H-1)k^{-\gamma},
\label{eq:FGN-autocorrel}
\end{equation}
where $\sum_{k=1}^{\infty} \rho(k) \rightarrow \infty$, when $k \rightarrow \infty$.  
For the FGN, the exponent of auto-correlation is defined by the Hurst parameter $\gamma = 2-2H$. We see that FBM is an essential process with various statistical properties defined by the Hurst parameter. Thus, researchers use an extensive choice of statistical estimators to determine $H$ and evaluate memory effects even when FBM or FGN are not applicable for modeling sample data. Let us introduce a generalization of the fractional processes essential for modeling social systems, where power-law noise can appear in the very origin of the process. One must decide which model to apply in the description of empirical data exhibiting anomalous diffusion $d \neq 0$, as observed dynamics can originate from the long-range memory or power-law of the noise.

\subsection{Fractional L\'{e}vy stable motion}
\label{sec:FLSM}
Gaussian distribution and models related to it are very important for the modeling of processes in nature. Nevertheless, the complexity of processes in social systems often leads to the power-law behavior, and the distributions with fat tails are just indispensable. Thus one needs to generalize the FBM to the case of non-Gaussian distributions. There is a special class of stable, invariant under summation, distributions \cite{Smarodinsky1994ChapmanHall}, useful in the modeling both super and sub-diffusion \cite{Klafter2012WorldScientific}. Here we are interested in the symmetric zero mean, stable distribution defined by the stability index $0<\alpha<2$. This new parameter generalizing the previous FBM process and defining the shape of the new distribution is responsible for the power-law tails of the new PDF $P(x) \sim \vert x \vert^{-1-\alpha}$. The new process called fractional L\'evy stable motion (FLSM), $L_H^{\alpha}(t)$, is $H$ self-similar and $\alpha$-stable with stationary increments. When $\alpha=2$ one can recover FBM discussed in previous subsection, see \cite{Smarodinsky1994ChapmanHall,Klafter2012WorldScientific} for the precise FLSM definition and relation with FBM.

Widely accepted and investigated FBM forms the background for the many estimators of Hurst exponent $H$. Accepting more general FLSM approach one has to reevaluate previously used methods \cite{Gontis2020JStatMech,Kazakevicius2021Entropy}, as we now have at least two independent parameters: the stability index $0<\alpha<2$ and memory parameter $d$. Since in the L\'evy stable case the second moment is infinite the measure of noise auto-correlation, e.g., the co-difference \cite{Smarodinsky1994ChapmanHall,Weron2005PRE}, is used instead of covariance
\begin{equation}
\tau(k) = \sim k^{-(\alpha-\alpha H)}.
\label{eq:co-difference}
\end{equation}   
We have to admit that the parameter $\gamma = \alpha-\alpha H = \alpha-\alpha d-1$, has the strong dependence on $\alpha$, when for the Gaussian processes it was considered just as the indicator of long-range memory. The sample MSD, Eq. \eqref{eq:sample-MSD}, should be a good estimator of memory parameter $d$, $M_N(k) \sim k^{(2d+1)}$, as stated in \cite{Burnecki2010PRE}. 

It is widely accepted to analyze stochastic processes using the PSD calculated by Fourier transform of sample time series and finding the ensemble average. Even for the FBM the interpretation of sample PSD is challenging as for $H>1/2$ PSD exhibits the same frequency dependence as Brownian motion, and this property may lead to false conclusions \cite{Krapf2019PRE}. Markov processes exhibiting PSD $S(f) \sim 1/f^{\beta}$ confirm that power-law properties of time series  contribute to the behavior of $S(f)$ \cite{Gontis2004PhysA,McCauley2006PhysA,McCauley2007PhysA,Micciche2009PRE,Micciche2013FNL,
Ruseckas2011PRE,Kazakevicius2015ChSF}. Thus, the extension of sample PSD analysis for FLSM is necessary to understand the relevant systems' behavior fully, and we do not include the PSD analysis in this contribution. 

Rescaled range analysis (R/S) \cite{Hurst1951,Beran1994Chapman,Montanari1999MCM} is one of the most popular estimators of $H$. The method relies on the measure of scaled fluctuations. Thus, first of all, it reveals the property of self-similarity. We have to be careful using this method to evaluate the long-range memory property as the relation with MSD $d=H-1/2$ and auto-correlation of increments $\gamma = 2-2H$ is valid only for the FBM. For the other processes with stable distribution, this method does not work properly as it involves normalization by the standard deviation \cite{Beran1994Chapman} thus, we do not use this method in this contribution. 

One more method to quantify scaled fluctuations in the sample time series is the multifractal detrended fluctuation analysis (MF-DFA) \cite{Peng1994PRE,Kantelhardt2002PhysA}. In the estimator's multifractal version, one can discriminate between fractal and multifractal behavior of sample time series. The method relies on the sample variance calculation; thus, it is not suitable for the non-Gaussian stable processes. 

Earlier, we have proposed the burst and inter-burst duration analysis as one more method to quantify the long-range memory through the evaluation of $H$ \cite{Gontis2017PhysA,Gontis2017Entropy,Gontis2018PhysA,Gontis2020JStatMech}. When one dimensional sample time series fluctuations are bounded, then any threshold divides these series into sequence of burst $T_j^b$ and inter-burst $T_j^i$ duration, $j=1,..N_b$. The notion of burst and inter-burst duration is directly related to the threshold first passage problem from the nearest its vicinity. The burst duration is the first passage time from above and inter-burst from below of the threshold, see \cite{Gontis2020JStatMech,Gontis2018PhysA,Gontis2017PhysA,Gontis2017Entropy} for more details. We have to revise the concept of burst duration analysis (BDA) from the more general perspective of FLSM.

Fortunately, Ding and Yang in \cite{Ding1995PhysRevE} demonstrated that PDF of burst or inter-burst duration $P(T)$ scales in some region as
\begin{equation}
P(T)\sim T^{H-2}.
\label{eq:T-PDF}
\end{equation} 
This property has to be applicable not only for the FBM but also for the other fractional processes with the non-Gaussian distribution. The authors demonstrated this scaling property for the deterministic chaotic processes, where the increments are not Gaussian and for the on-off intermittency with fractional noise modulation. Thus we do expect that the scaling \eqref{eq:T-PDF} might be applicable for the FLSM as well. Note that this scaling property is invariant regarding non-linear transformations of the time series \cite{Gontis2012ACS}, thus the self-similarity $H$ estimator based on this property might be applicable for the bounded diffusion cases. Nevertheless, from the perspective of more general FLSM processes, the question of which properties can be recovered using this method is open and has to be investigated. 

The method of Absolute Value estimator (AVE) has been considered as working correctly even for the time series with infinite variance \cite{Taqqu1995Fractals,Mercik2003APP,Weron2005PRE,Magdziarz2013JPhysA}. The method is based on mean value $\delta_n$ calculated from sample series $Y_i$ and evaluating its scaling with length of sub-series $n$. Sub-series $Y_j^{(n)}=\frac{1}{n} \sum_{i=(j-1)n+1}^{jn} Y_i$ divide original series into $m$ pieces of equal length $n$, $m * n=N$. Calculate $\delta_n$
\begin{equation}
\delta_n = \frac{1}{m} \sum_{j=1}^m \vert Y_j^{(n)}-\langle Y \rangle\vert,
\label{eq:AVE}
\end{equation} 
where $\langle Y \rangle$ is the overall series mean. Then the  absolute value scaling parameter $H_{AV}$ can be evaluated from the scaling relation
\begin{equation}
\delta_n \sim n^{H_{AV}-1}.
\label{eq:AVE-scaling}
\end{equation}

One more almost equivalent estimator of scaling properties of the time series is Higuchi's method \cite{Higuchi1988PhysD,Taqqu1995Fractals}. It relies on finding fractional dimension $D$ of the length of the path. The normalized path length $L_n$ in this method is defined as follows
\begin{equation}
L_n = \frac{N-1}{n^3} \sum_{i=1}^n \frac{1}{m-1} \sum_{j=1}^{m-1}\vert X_{i+jn}-X_{i+(j-1)n}\vert,
\label{eq:Higuchi}
\end{equation}
and $L_n \sim n^{-D}$, where $D=2-H$. We will use both methods, AVE and Higuchi's, for the empirical analysis in this contribution.

\section{Data \label{sec:data}}
In this contribution, we continue our efforts understanding the nature of the long-range memory phenomenon in socioeconomic systems \cite{Kazakevicius2021Entropy,Gontis2020JStatMech}. We aim to illustrate a numerical and heuristic approach to the sample empirical data of the financial market as the order flow data provide the best available self-similar time series of social systems with expected long-range memory property \cite{Lillo2004SNDE,Bouchaud2004QF,Toth2015JEDC}. Our source of empirical data is limit order book data for all NASDAQ traded stocks provided by Limit Order Book System LOBSTER \cite{Huang2011Lobster}. The limit order book (LOB) data that LOBSTER reconstructs originates from NASDAQ's Historical TotalView-ITCH files (http://nasdaqtrader.com). Here we extend the list of stocks considered in \cite{Gontis2020JStatMech} and construct daily time series of order flow in the period from 3 to 31 of August in 2020, a total of 21 working days. We use the sample data from other periods to show that the main statistical properties we are interested in probably are independent of the selected period.  

\begin{figure}
\begin{centering}
\includegraphics[width=0.9\textwidth]{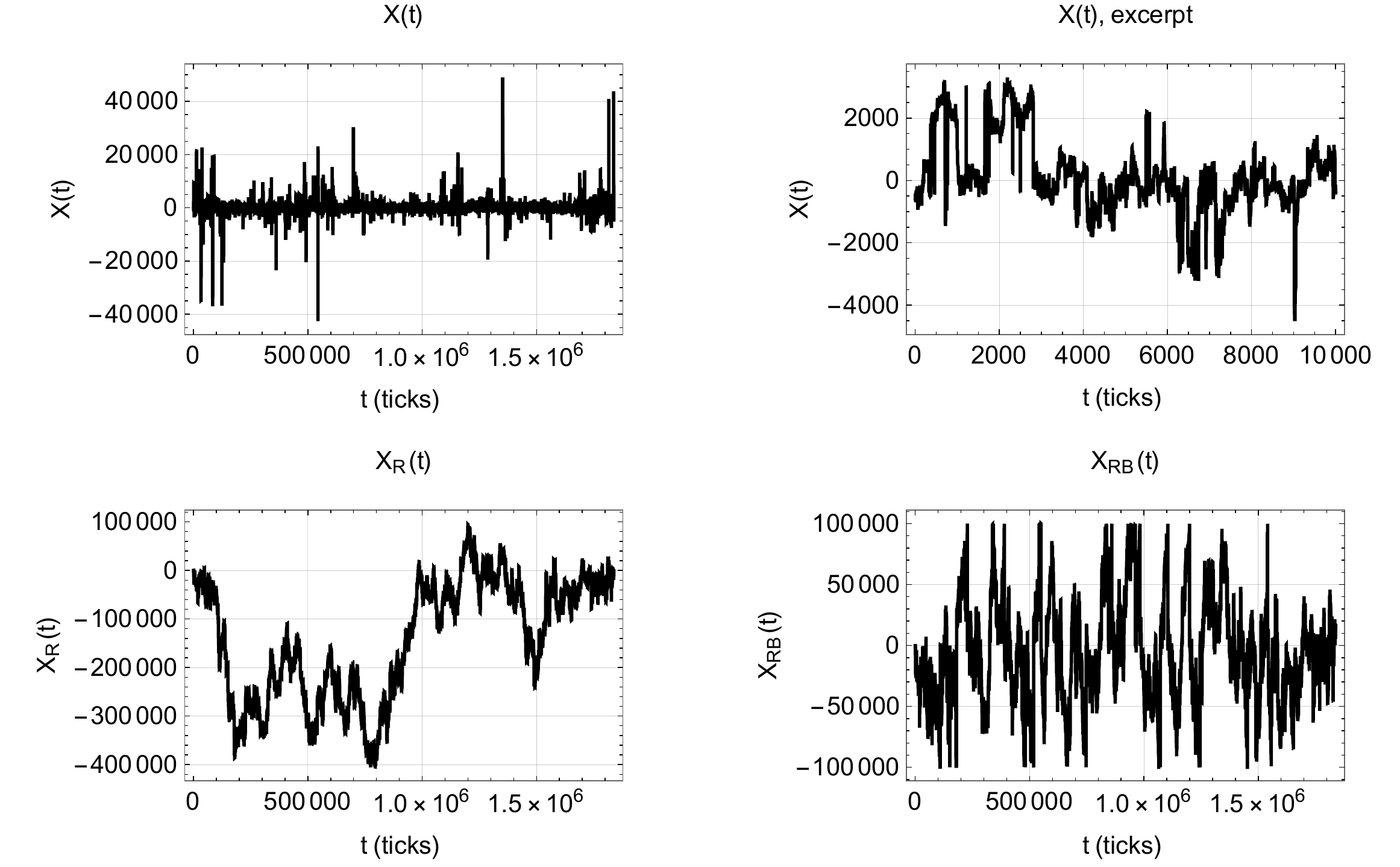}
\par\end{centering}
\caption{Examples of AAPL stock order disbalance tick time series in the one trading day period. $X(t)$ denotes empirical time series as defined by Eq. \eqref{eq:order-disbalance}; its excerpt is just the same series of $10000$ ticks duration, $X_R(t)$ denotes the same series, when increments are randomized; and $X_{RB}(t)$ denotes randomized and bounded series. \label{fig1}}
\end{figure}

We use LOBSTER data files: message.csv and orderbook.csv for each selected trading day and ticker (stock). The message.csv file contains the full list of events causing an update of LOB in the selected price range. We investigate orders up to the ten levels of prices in this research. Both files provide exact information about the instantaneous state of LOB needed to define order disbalance time series. Any event $j$ changing the LOB state has a time value $t_j$ and order book has full list of volumes for the 10 price levels: 10 buy volumes $v_k^{+}(t_j)$ as well as 10 sell volumes $v_k^{-}(t_j)$. Seeking for the most simple interpretation of results, in this contribution, we define order disbalance event time series $X(j)$ as the difference between all buy and sell orders or simple sum of disbalance increments $Y(i)$
\begin{equation}
X(j)=\sum_{k=1}^{10}(v_k^{+}(j)-v_k^{-}(j))=V^{+}(j)-V^{-}(j)=\sum_{i=1}^{j} Y(i).
\label{eq:order-disbalance}
\end{equation}
Compare it with the normalized version used in \cite{Gontis2020JStatMech} $X_N(j)=(V^{+}(j)-V^{-}(j))/(V^{+}(j)+V^{-}(j))$, where $-1<X_N(j)<1$. The simplified definition of disbalance is preferable as it helps to establish a relationship with the discrete sum of random volume flow $Y(i)$ and a particular case of ARFIMA process \cite{Burnecki2017ChaosSF}. The first two sub-figures in Fig. \ref{fig1} illustrate the example of simplified order disbalance empirical time series for the AAPL stock. The considered simplified empirical order disbalance time series still can serve as an example of opinion dynamics in the social system.

We will show below that PDF of $Y(i)$ has symmetric power-law tails; thus, the time series $X(j)$ might behave similarly as FLSM or a particular case of ARFIMA process. Consequently, scaling properties of these series depend on PDF characterized by $\alpha$ and memory characterized by $d$ of increment series $Y(i)$. To clarify the different impacts of $\alpha$ and $d$ on the estimated scaling parameters, we construct an artificial time series $X_R(j)$ excluding memory effects. The proposed method to manipulate original empirical data helps quantitatively estimate  the impacts of $\alpha$ and $d$. Thus, by replacing sequence of $Y(i)$ values in Eq. \eqref{eq:order-disbalance} by the random (reshuffled) sequence of the same empirical values $Y_R(i)=Random[Y(i)]$ we construct an artificial time series $X_R(j)$ with $d=0$. One can observe in Fig. \ref{fig1} that though the series $X(t)$ look bounded, in the case of random increments, series $X_R(i)$ become unbounded and non-stationary. AVE and Higuchi's methods work correctly with non-stationary time series. Unbounded series is a problem for the burst duration analysis, as they have only a few intersections with the threshold. We will need one more transformation of the series $X_R(j)$ introducing bounds of diffusion $\vert X_{RB}(j) \vert \leq 100000$, see example in Fig. \ref{fig1}. This diffusion restriction is introduced as the soft boundary condition $X_{RB}(i+1) = \max(\min(X_{RB}(i)+Y_R(i), 100000), -100000)$. Note that the bounded random series $X_{RB}(i)$ must have the same exponent of inter-burst duration PDF scaling as this property is invariant regarding such transformation \cite{Ding1995PhysRevE,Gontis2020JStatMech}.

\section{Results \label{sec:reults}}
We consider the simplified order disbalance time series retrieved from LOBSTER data as described in the previous section. We seek to reveal the behavior of the Hurst parameter estimators regarding shuffling of the empirical series increments. The procedure of increment shuffling is aimed at the task to compare the impacts of increment PDF and increment auto-correlation. The PDF of order volumes or, more specifically, in this research of order disbalance increments $Y(i)$ is rather sophisticated with resonance structure, power-law tails, and some variability from stock to stock. Nevertheless, this PDF is highly symmetric and stable from day to day, thus contributes to the scaling properties of the time series constructed from the flow of order volumes. 

\begin{figure}
\begin{centering}
\includegraphics[width=0.95\textwidth]{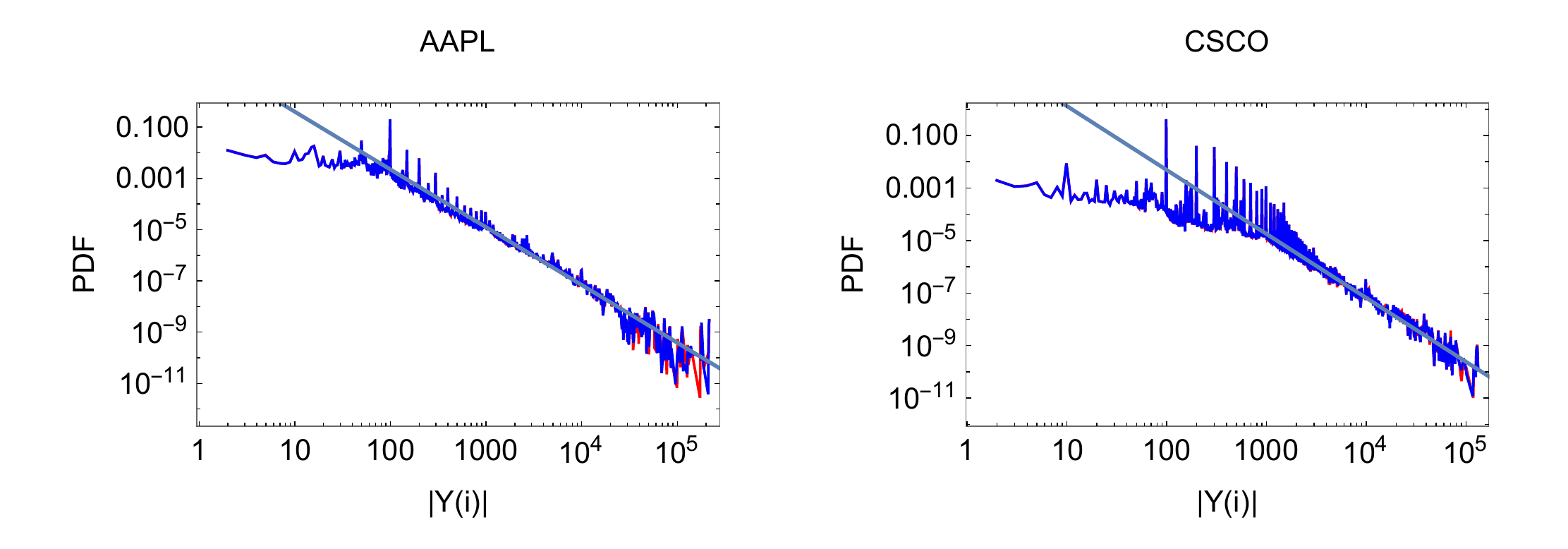}
\par\end{centering}
\caption{Histograms of increment $\vert Y(i) \vert$ positive (red) and negative (blue) values.
Histograms are calculated from the joint daily series in the whole period of 21 day for the two stocks: AAPL, and  CSCO. Straight lines fit the PDF tails on the log-log scale, exponent values: $\nu_1=2.25$ for  AAPL and $\nu_2=2.43$ for CSCO. \label{fig2}}
\end{figure}

 In figure \ref{fig2} we illustrate histograms of increment absolute values, $\vert Y(i) \vert$,  calculated for the two stocks: AAPL, CSCO. These histograms are calculated from the joint daily series in the whole period of 21 days. Two collar curves denoting PDFs of positive and negative values are almost indispensable, but power-law fitting straight lines give different exponent values: $\nu_1=2.25$ for  AAPL and $\nu_2=2.43$ for CSCO. Note the resonance structure observed for the round numbers of volumes.  

\begin{figure}
\begin{centering}
\includegraphics[width=0.95\textwidth]{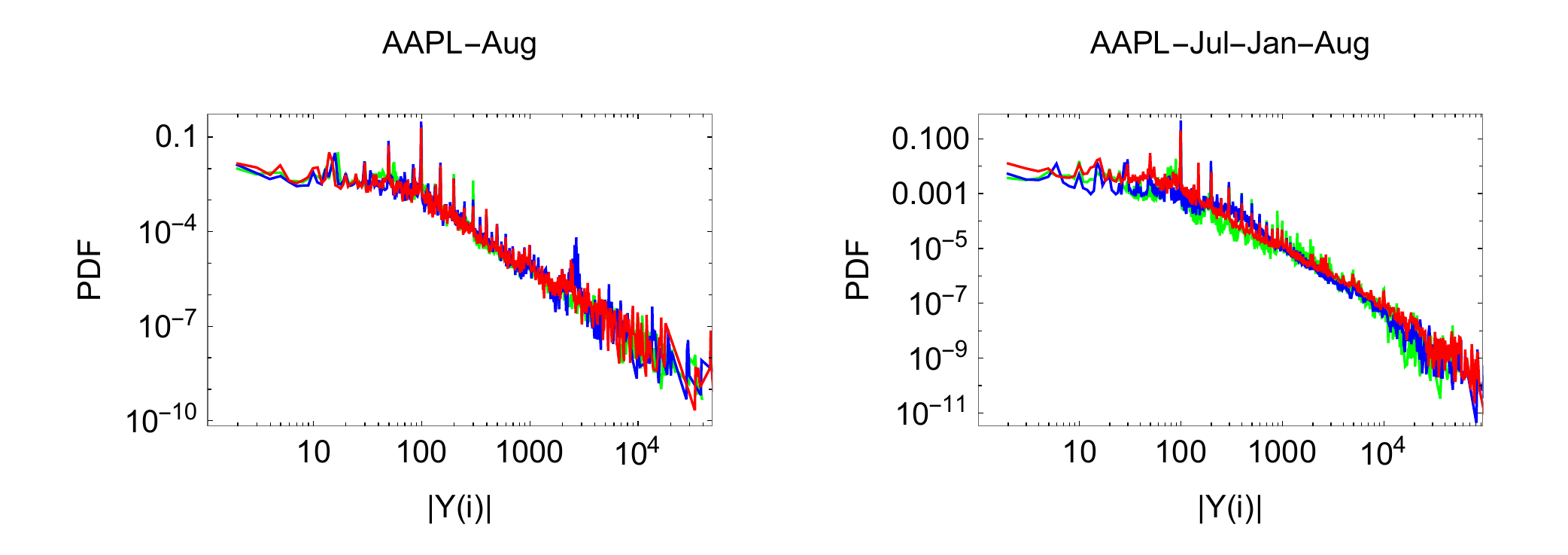}
\par\end{centering}
\caption{Histograms of AAPL disbalance increment $\vert Y(i) \vert$ calculated for different periods. Histograms in the left sub-figure are calculated for 3 different trading 
days: (green) in 03.08.2020; (blue) in 14.08.2020; (red) in 28.08.2020. Histograms in right sub-figure for 3 different trading months: (green) July of 2012; (blue) January of 2020; (red) August of 2020. \label{fig3}}
\end{figure}

The non-Gaussian PDFs of increments are essential characteristics of stocks investigated from the perspective of FLSM. Thus, it is useful to know whether PDFs are the same in the various periods of observation. In figure \ref{fig3} we compare daily and monthly histograms of AAPL disbalance increments $\vert Y(i) \vert$ calculated for the distant periods. The presented example and even more detailed investigation confirm that PDFs of increments are essential characteristics of stocks considered. Nevertheless, we admit that deviations from the most probable form of PDF can appear in the specific trading periods, but we leave these peculiarities outside the scope of this research. 

\begin{figure}
\begin{centering}
\includegraphics[width=0.95\textwidth]{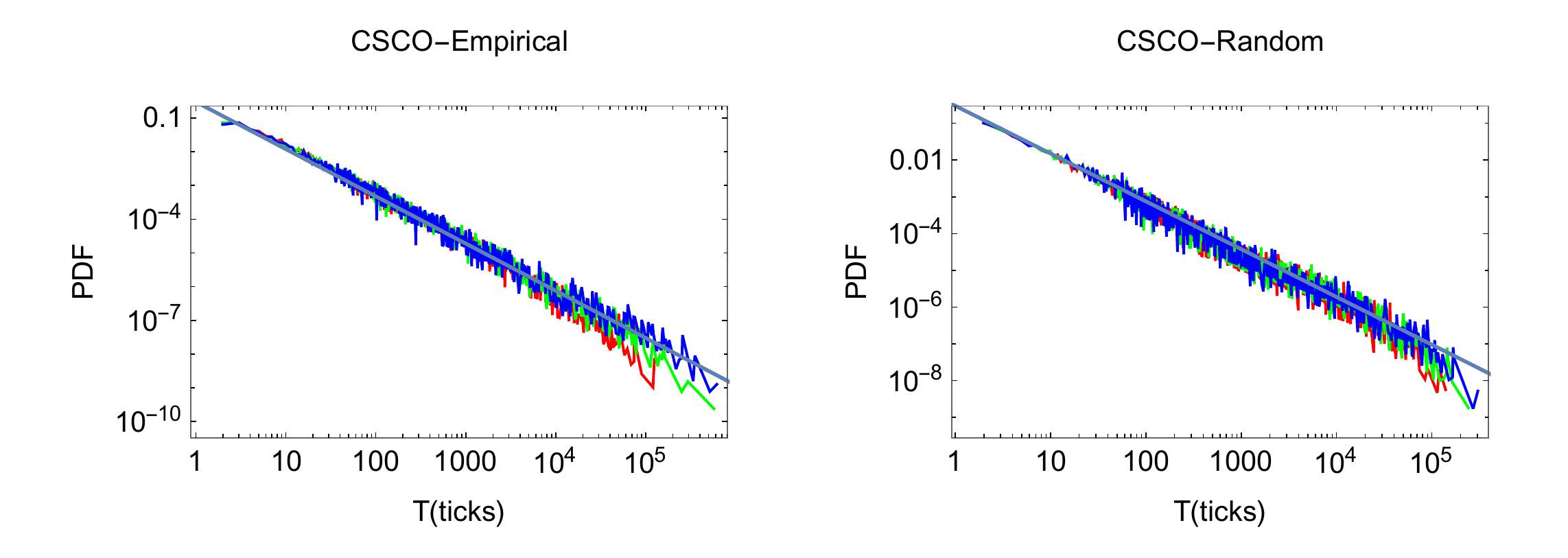}
\par\end{centering}
\caption{PDF of inter-burst duration $T$  for the CSCO order disbalance event time series.
Sub-figures show: CSCO-Empirical PDF is calculated for the joint daily series of the whole period of 21 trading day with three values of threshold $\{0.5,1.0,1.5\}*\sigma$; CSCO-Random PDF is calculated from the same empirical series using random reshuffling  of $Y(i)$ and bounding procedure $X_{RB}(j)$, with the same values of threshold. The plots are red for the lowest thresholds, green for the higher values and blue for the highest. Straight lines are power-laws fits with exponents $\eta=1.4$ for the empirical series, and $\eta_R=1.3$ for random. \label{fig4}}
\end{figure}

The interpretation of burst duration analysis with the assumption of FLSM is a challenging
task. It is well known that the first-passage time PDF for any one-dimensional Markovian process with symmetric jump length distribution, including L\'evy flights, has the universal Sparre Andersen asymptotic $P(T) \sim T^{3/2}$ \cite{Andersen1953MathSc,Andersen1954MathSc,Metzler2014Springer,Padash2019JPhysA,Palyulin2019NJPhysics}. 
Nevertheless, in the FLSM case, when $d=0$ and $H=1/\alpha$, one might expect the first-passage time PDF $P(T) \sim T^{2-1/\alpha}$, where $\alpha=\nu-1$, different from Sparre Andersen asymptotic of $3/2$. We analyze inter-burst duration PDF for the empirical order disbalance time series and for the shuffled time series of various stocks; see the examples of CSCO inter-burst duration histograms in the figure \ref{fig4}.
Both histograms exhibit power-law behavior in the region up to the five orders of $T$ values.
In the case of random order flow, the PDFs of inter-burst duration $T_R$ for different threshold values $\{0.5,1.0,1.5\}*\sigma$ almost coincide, where  $\sigma$ denotes the standard deviation of time series in the period of investigation. In empirical time series, one observes the cutoff of power-law for the high $T$ values. The area of power-law is wider for the higher threshold values. 

Very preliminary BDA of discrete FLSM series generated using ARFIMA\{0,d,0\} process with $d\neq 0$ revealed that burst and inter-burst duration PDFs depend on the threshold values giving different estimates of power-law exponents $\eta$. Thus, we choose in this contribution to deal with zero thresholds when the burst and inter-burst duration PDFs of symmetric time series coincide. We consider eighteen stocks, plot PDFs of burst duration for empirical and random series, fit the power-law part for the lowest values of duration $T$, and define fitting exponents $\eta$ for the empirical series and $\eta_R$ with random increments. Then corresponding Hurst exponents are evaluated using Eq. \eqref{eq:T-PDF}: $H_{BD}=2-\eta$, and $H_{BDR}=2-\eta_R$. In the table \ref{table1} we provide our results of the evaluated scaling exponents using estimators
described in Section \ref{sec:FLSM} and evaluated using BDA.
\begin{table}
\centering
\caption{
Scaling exponents of the order disbalance time series for the 18 stocks. Stocks are listed in the first column. Evaluated exponents are listed in the first row as follows: $\lambda$ is exponent of MSD Eq. \eqref{eq:sample-MSD}; $H_{AV}$ is $H$ evaluated using AVE; $H_{AVR}$ is the same for the randomized series; $H_{Hig}$ is $H$ evaluated using Higuchi's method; $H_{HigR}$ is the same for the randomized series; $H_{BD}$ is $H$ evaluated using BDA; $H_{BDR}$ is the same for the randomized series; and $1/\alpha$ is one over $\alpha$, exponent of disbalance increment PDF.}
\begin{tabular}{|l+l|l|l|l|l|l|l|l|l|l|}
\hline
\multicolumn{1}{|l|}{\bf Stock} & \multicolumn{1}{|l|}{\bf $\lambda$} & \multicolumn{1}{|l|}{\bf $H_{AV}$} & \multicolumn{1}{|l|}{\bf $H_{AVR}$} & \multicolumn{1}{|l|}{\bf $H_{Hig}$} & \multicolumn{1}{|l|}{\bf $H_{HigR}$} & \multicolumn{1}{|l|}{\bf $H_{BD}$} & \multicolumn{1}{|l|}{\bf $H_{BDR}$} & \multicolumn{1}{|l|}{\bf $1/\alpha$}\\ \thickhline
AAPL & $0.65$ & $0.27$ & $0.63$ & $0.30$ & $0.64$ & $0.47$ & $0.75$ & $0.80$\\ \hline
ABBV & $0.84$ & $0.25$ & $0.53$ & $0.26$ & $0.53$ & $0.46$ & $0.57$ & $0.38$\\ \hline
ABT  & $0.74$ & $0.16$ & $0.51$ & $0.17$ & $0.50$ & $0.45$ & $0.54$ & $0.44$\\ \hline
ACN  & $0.65$ & $0.12$ & $0.50$ & $0.15$ & $0.50$ & $0.34$ & $0.53$ & $0.35$\\ \hline
ADI  & $0.71$ & $0.19$ & $0.50$ & $0.18$ & $0.51$ & $0.44$ & $0.67$ & $0.40$\\ \hline
ADP  & $0.69$ & $0.16$ & $0.51$ & $0.16$ & $0.51$ & $0.45$ & $0.58$ & $0.41$\\ \hline
ADBE & $0.59$ & $0.20$ & $0.51$ & $0.20$ & $0.52$ & $0.45$ & $0.63$ & $0.62$\\ \hline
ADSK & $0.57$ & $0.20$ & $0.52$ & $0.20$ & $0.53$ & $0.38$ & $0.62$ & $0.62$\\ \hline
AMD  & $0.84$ & $0.31$ & $0.60$ & $0.33$ & $0.61$ & $0.35$ & $0.64$ & $0.65$\\ \hline
AMZN & $0.52$ & $0.23$ & $0.54$ & $0.24$ & $0.55$ & $0.40$ & $0.63$ & $0.67$\\ \hline
ASML & $0.47$ & $0.14$ & $0.51$ & $0.15$ & $0.51$ & $0.50$ & $0.58$ & $0.39$\\ \hline
AVGO & $0.55$ & $0.18$ & $0.56$ & $0.23$ & $0.56$ & $0.41$ & $0.65$ & $0.67$\\ \hline
BABA & $0.62$ & $0.21$ & $0.57$ & $0.26$ & $0.59$ & $0.49$ & $0.71$ & $0.74$\\ \hline
CSCO & $0.91$ & $0.34$ & $0.56$ & $0.36$ & $0.56$ & $0.52$ & $0.62$ & $0.70$\\ \hline
DIS  & $0.75$ & $0.21$ & $0.54$ & $0.25$ & $0.55$ & $0.41$ & $0.61$ & $0.56$\\ \hline
FB   & $0.64$ & $0.24$ & $0.62$ & $0.27$ & $0.63$ & $0.45$ & $0.75$ & $0.76$\\ \hline
GOOG & $0.40$ & $0.20$ & $0.54$ & $0.21$ & $0.54$ & $0.31$ & $0.60$ & $0.59$\\ \hline
INTC & $0.91$ & $0.37$ & $0.59$ & $0.40$ & $0.60$ & $0.51$ & $0.66$ & $0.72$\\ \hline
\end{tabular}
\label{table1}
\end{table}

For the eighteen stocks listed, we provide eight exponents. In the first column of numbers, we give exponents $\lambda$ of sample MSD, Eq. \eqref{eq:sample-MSD}, for the empirical order disbalance series. We do not list exponent $\lambda$ of MSD for the random series as the all values are as expected $\lambda=1$.  The evaluated MSD gives us the average values of the memory parameter $d_{MSD}=-0.16$ with a standard deviation $0.08$ exhibiting the behavior of sub-diffusion. One can observe considerable fluctuations of MSD from stock to stock, indicating variation of the memory effects or the sensitivity of the estimator to the stock dependent peculiarities of time series.

Absolute value estimator, Eq. \eqref{eq:AVE}, see columns $H_{AV}$, and $H_{AVR}$, gives considerably different values of the Hurst parameter for the empirical and randomized time series. With the assumption of FLSM $H=d+1/\alpha$ we can evaluate $d$ as $d=H_{AV}-H_{AVR}$. We get the mean value for the stocks considered $d_{AV}=-0.32$ and standard deviation $0.05$. In columns $H_{Hig},\quad H_{HigR}$, we provide results of evaluated $H$ using Higuchi's estimator Eq. \eqref{eq:Higuchi}. Both methods give very similar results; we get mean value of $d$ evaluated by Higuchi's estimator $d_{Hig}=-0.31$ and the same standard deviation $0.05$. The shift from empirical to randomized sequence of increments $Y_R(i)$ gives us a more stable estimation of $d$ than straightforward use of sample MSD, Eq. \eqref{eq:sample-MSD}. From our perspective, this might be related to the impact of the bounded nature of empirical time series and considerable deviations of the empirical increment PDFs from L\'evy stable form.

In the columns $H_{BD},\quad H_{BDR}$, we list $H$ values defined fitting histograms of burst duration for the empirical and randomized series with zero threshold. The values are scattered in the interval $H=0.31\div 0.52$ for the empirical and in the interval $H=0.53\div 0.75$ for the random series. From the difference of $H$ for the empirical and randomized series, we get the mean value for the stocks considered $d_{BD}=-0.20$ and the standard deviation $0.07$. The estimated memory parameter is closer to the value we get from the MSD method. Values $1/\alpha$ in the last column, defined from the power-law of $Y(i)$ PDF, see example in Fig. \ref{fig2}, are scattered in the wide interval $0.35 \div 0.80$. $\alpha=\nu-1$, where $\nu$ denotes the exponent of power-law tail of $Y(i)$ PDF. Note that empirical PDFs of $Y(i)$ considerably deviate from the form of stable distributions; thus, one can observe values of $1/\alpha$ smaller than $1/2$. We do not use these empirical values of $\alpha$ in evaluating $H$. Instead, the method assumes the memory parameter as $H$ difference for empirical and randomized series $d=H-H_R$. Likely, such an assumption might work even when deviations from L\'evy stable form are present. The observed stability of defined memory parameters for various stocks probably supports this assumption.

\begin{figure}
\begin{centering}
\includegraphics[width=0.9\textwidth]{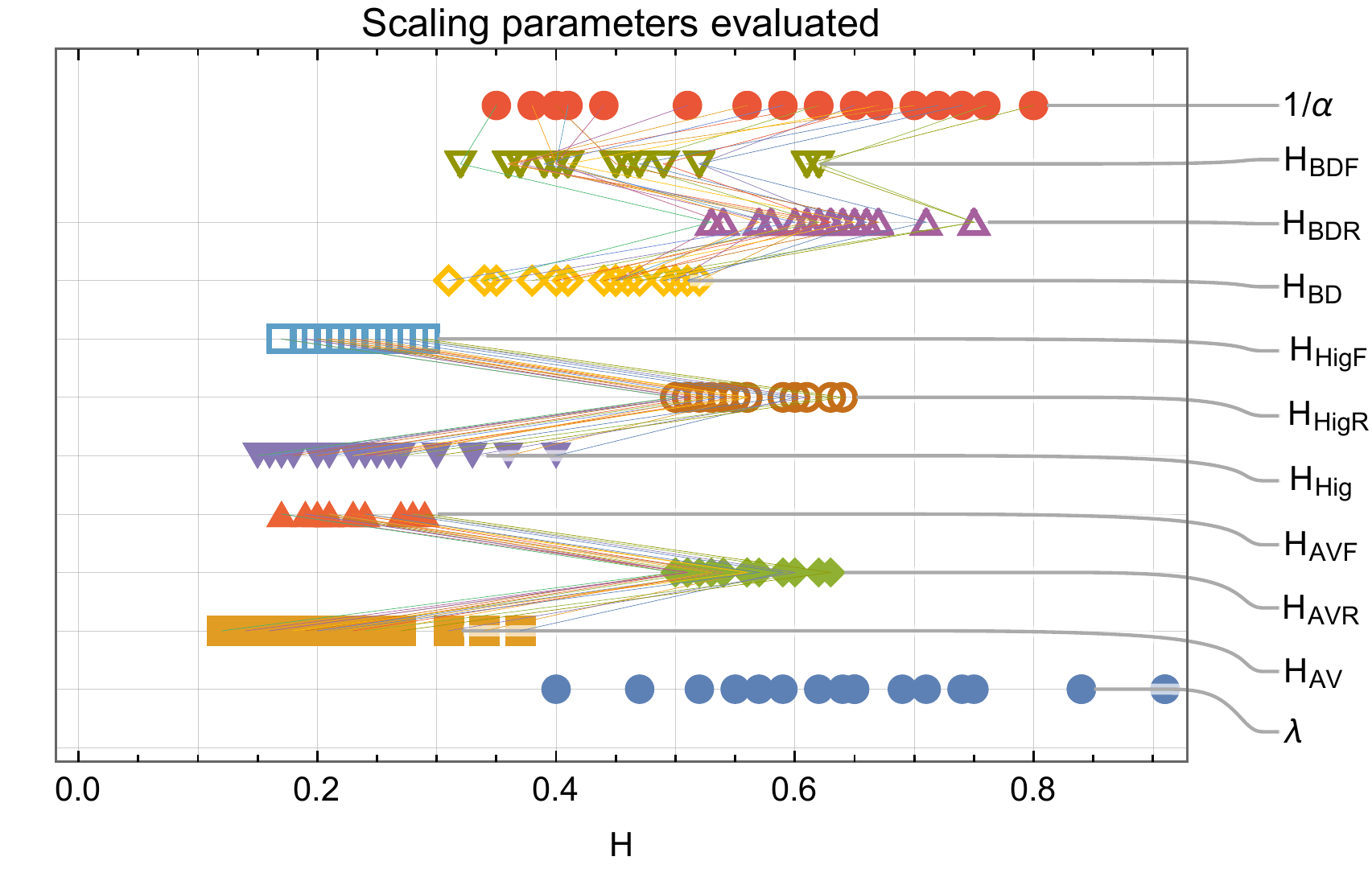}
\par\end{centering}
\caption{Comparison of Hurst and other scaling exponents defined for the empirical and randomized order disbalance time series. All rows have 18 values corresponding to the stocks investigated. Scaling parameters are labeled as defined in the text and caption of the table \ref{table1}. Pairs of values for the same stock: $H_{AV}$ and $H_{AVR}$; $H_{AVR}$ and $H_{AVF}$; $H_{Hig}$ and $H_{HigR}$; $H_{HigR}$ and $H_{HigF}$; $H_{BD}$ and $H_{BDR}$; $H_{BDR}$ and $H_{BDF}$; $H_{BDF}$ and $1/\alpha$ are joined by straight lines to reveal the impact of memory effects.  \label{fig5}}
\end{figure}

To get a better sense of various scaling exponents provided in the table \ref{table1} we present them in Fig. \ref{fig5}. The pairs of values for the same stock: $H_{AV}$ and $H_{AVR}$; $H_{Hig}$ and $H_{HigR}$; $H_{BD}$ and $H_{BDR}$ are joined by straight lines to reveal the impact of memory effect. $H$ estimates: BDR, HigR, AVR, give us less scattered values of $H$ for various stocks than $1/\alpha$ as one should expect from the theory of FLSM \cite{Smarodinsky1994ChapmanHall,Weron2005PRE}. We have to admit here that empirical PDFs of increments have considerable deviations from the theoretical stable distributions with parameter $\alpha$. 
Note that $H$ shifts of the randomized series are very stable for various stocks, as lines in Fig. \ref{fig5} indicate. Nevertheless, the shift in the BDA case is smaller and gives the estimation of $d_{BD}$ considerably closer to the MSD estimation. Probably, both methods give lower and more scattered values of $d_{MSD}$ indicating the dependence of the methods on the $\alpha$ and other stock-dependent peculiarities of the time series.
 From our perspective, this might be related to the bounded nature of empirical order disbalance series as well.
\begin{figure}
\begin{centering}
\includegraphics[width=0.95\textwidth]{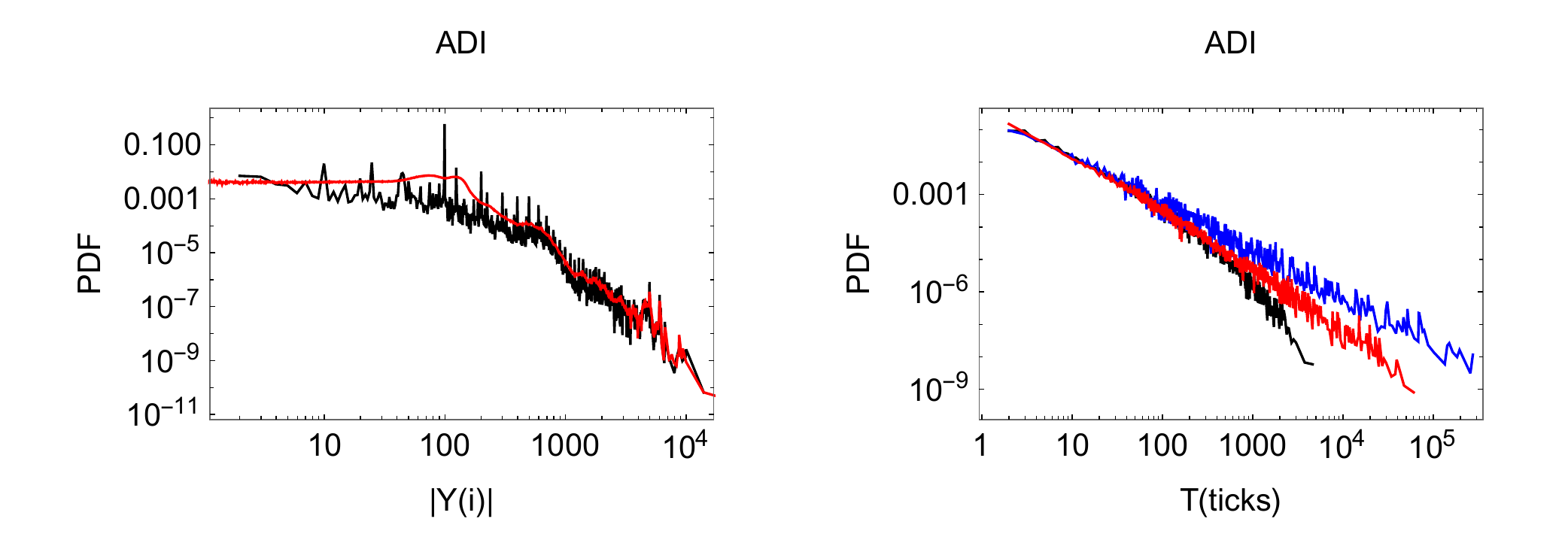}
\par\end{centering}
\caption{Comparison of increment PDFs and inter-burst duration PDFs of the fractionally reverted order disbalance series for the ADI stock. Left sub-figure exhibits the histogram of increments $\vert Y(i) \vert$ for the original empirical time series, black line, and the histogram of increments $\vert Z_R(i) \vert$ for the reverted fractional time series, red line. 
Histograms are calculated from the joint daily series in the whole period of 21 days. Right sub-figure exhibits: (black) burst duration PDF for the original empirical order disbalance time series $X(i)$; (blue) the PDF of randomized time series $X_R(i)$; (red) the PDF of fractionally reverted time series $X_F(i)$. \label{fig6}}
\end{figure}

We have an indirect opportunity to check whether the evaluated value of memory parameter $d \simeq -0.3$ defined using Higuchi's and Absolute value estimators is reasonable. One can get back to the correlated time series using the standard fractional sum procedure. We denote the random sequence of increments as $Z(i)=Random[Y(i)]$, than we can get back to the anti-correlated time series $Y_F(i)$ using the most simple version of the ARFIMA(0,d,0) process  \cite{Stoev2004Fractals,Burnecki2017ChaosSF}. Note that the
both acronyms 'ARFIMA' and 'FARIMA' are used interchangeably in the literature.
Let us consider that increments $Y_F(i)$ are just the fractional sums
\begin{equation}
X_F(i)= X_F(i-1) + \sum_{j=0}^{\infty} \frac{\Gamma(j+d)}{\Gamma(d) \Gamma(j+1)} Z(i-j) = X_F(i-1) + Y_F(i),
\label{eq:ARFIMA}
\end{equation}
where $\Gamma$ is the Gamma function. In the left sub-figure of Fig. \ref{fig6} we compare PDFs of original ADI order disbalance increments $\vert Y(i) \vert$ with reverted back increments $\vert Y_F(i) \vert$ using parameter $d=-0.3$ and Eq. \eqref{eq:ARFIMA}. The infinite sum in Eq. \eqref{eq:ARFIMA} we approximate with the sum over $1000$ terms. Both PDFs, see black line for $\vert Y(i) \vert$ and  red for $\vert Y_F(i) \vert$ almost coincide, only resonance structure was not recovered in the formal reversion procedure. 

It is useful to check out other estimators with new reverted time series $X_F(i)$. On the right sub-figure of Fig. \ref{fig6} we illustrate the PDFs of burst duration $T$, calculated for the time series: $X(i)$, $X_R(i)$, $X_F(i)$ of ADI stock. Though there are many different peculiarities of the original $X(i)$ and reverted $X_F(i)$ time series, the correspondence of scaling exponents is pretty good, see table \ref{table2} and Fig. \ref{fig5} for the comparison of scaling exponents for the original and reverted time series of eighteen stocks. Note that exponents $\lambda_F$ of MSD for the all reverted time series coincide, and the corresponding value $d_{MSDF}=-0.225$. Our numerical result given in figure $8$ of \cite{Kazakevicius2021Entropy} confirms theoretical prediction for the sample MSD $M_N(k) \sim k^{2d+1}$ \cite{Burnecki2010PRE} of accumulated ARFIMA\{0,d,0\}. This independence of MSD of stability parameter $\alpha$ probably explains the empirical observation independent of stock. AVE and Higuchi's methods estimate the memory parameter of reverted time series $d_{AVF}=d_{HigF}=-0.325$ with low standard deviation $0.01$. The increased precision of the evaluated memory $d$ of reverted series $X_F(i)$ using AV and Hig methods strengthens confidence that Absolute value and Higuchi's estimators are applicable even for the time series with power-law increment PDFs deviating from exact L\'evy stable form. 
\begin{table}
\centering
\caption{
Scaling exponents of the original and reverted order disbalance time series for the eighteen stocks. Stocks are listed in the first column and evaluated exponents in the first row, see used notations in the text and table \ref{table1}.}
\begin{tabular}{|l+l|l|l|l|l|l|l|l|l|l|}
\hline
\multicolumn{1}{|l|}{\bf Stock} & \multicolumn{1}{|l|}{$\lambda$} & \multicolumn{1}{|l|}{$\lambda_F$} & \multicolumn{1}{|l|}{$H_{AV}$} & \multicolumn{1}{|l|}{$H_{AVF}$} & \multicolumn{1}{|l|}{$H_{Hig}$} & \multicolumn{1}{|l|}{$H_{HigF}$} & \multicolumn{1}{|l|}{$H_{BD}$} & \multicolumn{1}{|l|}{$H_{BDF}$}\\ \thickhline
AAPL & $0.65$ & $0.55$ & $0.27$ & $0.29$ & $0.30$ & $0.29$ & $0.47$ & $0.61$  \\ \hline
ABBV & $0.84$ & $0.55$ & $0.25$ & $0.21$ & $0.26$ & $0.21$ & $0.46$ & $0.40$   \\ \hline
ABT  & $0.74$ & $0.55$ & $0.20$ & $0.17$ & $0.19$ & $0.17$ & $0.45$ & $0.40$   \\ \hline
ACN  & $0.65$ & $0.55$ & $0.12$ & $0.17$ & $0.15$ & $0.17$ & $0.34$ & $0.32$   \\ \hline
ADI  & $0.71$ & $0.55$ & $0.19$ & $0.20$ & $0.18$ & $0.20$ & $0.44$ & $0.45$   \\ \hline
ADP  & $0.69$ & $0.55$ & $0.16$ & $0.19$ & $0.16$ & $0.19$ & $0.45$ & $0.40$   \\ \hline
ADBE & $0.59$ & $0.55$ & $0.20$ & $0.21$ & $0.20$ & $0.21$ & $0.45$ & $0.36$   \\ \hline
ADSK & $0.57$ & $0.55$ & $0.20$ & $0.20$ & $0.20$ & $0.20$ & $0.38$ & $0.46$   \\ \hline
AMD  & $0.84$ & $0.55$ & $0.31$ & $0.28$ & $0.33$ & $0.28$ & $0.35$ & $0.52$   \\ \hline
AMZN & $0.52$ & $0.55$ & $0.23$ & $0.23$ & $0.24$ & $0.23$ & $0.40$ & $0.49$   \\ \hline
ASML & $0.47$ & $0.55$ & $0.14$ & $0.19$ & $0.15$ & $0.19$ & $0.50$ & $0.37$   \\ \hline
AVGO & $0.55$ & $0.55$ & $0.18$ & $0.23$ & $0.23$ & $0.23$ & $0.41$ & $0.41$   \\ \hline
BABA & $0.62$ & $0.55$ & $0.21$ & $0.24$ & $0.26$ & $0.25$ & $0.49$ & $0.52$   \\ \hline
CSCO & $0.91$ & $0.55$ & $0.34$ & $0.56$ & $0.36$ & $0.60$ & $0.52$ & $0.36$   \\ \hline
DIS  & $0.75$ & $0.55$ & $0.21$ & $0.21$ & $0.25$ & $0.22$ & $0.41$ & $0.36$   \\ \hline
FB   & $0.64$ & $0.55$ & $0.24$ & $0.28$ & $0.27$ & $0.28$ & $0.45$ & $0.62$   \\ \hline
GOOG & $0.40$ & $0.55$ & $0.20$ & $0.54$ & $0.21$ & $0.58$ & $0.31$ & $0.39$   \\ \hline
INTC & $0.91$ & $0.55$ & $0.37$ & $0.27$ & $0.40$ & $0.27$ & $0.51$ & $0.47$   \\ \hline
\end{tabular}
\label{table2}
\end{table}

Empirical analysis of order flow in the financial markets data from the perspective of FLSM or ARFIMA time series gives us a much more comprehensive understanding of this social system. The Absolute value and Higuchi's estimators of the Hurst parameter work exceptionally well with these time series despite considerable deviations of the noise from the exact L\'evy stable form. More extensive investigation of Burst duration analysis from FLSM or ARFIMA time series perspective and explanation of quantitative differences from the AVE and Higuchi's methods is needed.

\section{Conclusions}
Here we continue our efforts in understanding the long-range memory phenomenon in social systems \cite{Kazakevicius2021Entropy}. In previous work \cite{Gontis2020JStatMech} we investigated statistical properties of burst and inter-burst duration in order disbalance real time and event time series seeking to discriminate true and spurious long-range memory opportunities. These results, as previous findings in \cite{Lillo2004SNDE,Bouchaud2004QF,Toth2015JEDC} showed that the limit order flow exhibits strong positive autocorrelation. Here we admit that many widely used estimators of long-range memory were developed with the assumption of Gaussian noise distribution \cite{Taqqu1995Fractals} thus, the more general approach based on the FLSM or ARFIMA models has to be implemented for the order flow analysis in the financial markets, and other social systems \cite{Burnecki2010PRE,Burnecki2017ChaosSF}. More careful investigation of order flow sizes (tick sizes) revealed that PDFs of tick sizes are specific for each stock with some power-law tail. Exponents of power-law tails range from $2.25$ for AAPL to $3.86$ for ACN. Thus the order disbalance time series, as defined by Eq. \eqref{eq:order-disbalance}, is just like the most simple case of accumulated ARFIMA(0,d,0) process with empirical power-law PDF of noise deviating from the Gaussian and even from the stable L\'evy form. 

The empirical time series $X(j)$ with specific power-law increments $Y(i)$ behave like self-similar process. Many previously used estimators are not applicable in this case, as variance of $Y(i)$ and $X(j)$ might diverge in the continuous limit, see \cite{Beran1994Chapman,Taqqu1995Fractals}. We selected estimators of self-similarity in this contribution: Absolute Value and Higuchi's, which are applicable for the non-Gaussian case. We include Burst duration analysis as well, seeking a further investigation of possible applications of this method. The first finding of this research is that empirically defined Hurst parameters using Absolute value and Higuchi's estimators for all stocks is lower than $0.5$, compare with previous findings \cite{Lillo2004SNDE,Bouchaud2004QF,Toth2015JEDC,Gontis2020JStatMech} fluctuating around $0.7$. The previous results are more consistent with the values of the Hurst parameter we get with randomized time series $X_R(j)$, see $H_{AVR}$ and $H_{HigR}$ columns in the table \ref{table1} and figure \ref{fig5}. Nevertheless, the most meaningful result is that comparison of Hurst parameters for the empirical and randomized series gives us an estimation of memory parameter $d$, $d_{AV}=H_{AV}-H_{AVR}$ and $d_{Hig}=H_{Hif}-H_{HigR}$. Both estimators give almost the same average values: $d_{AV}=-0.32$, and $d_{Hig}=-0.31$, with the same standard deviation $0.05$ calculated for the list of randomly selected eighteen stocks. The result implies that order disbalance time series can be considered anti-persistent with a pretty stable $d$.

We implemented an indirect check of FLSM or ARFIMA process assumption by the procedure of randomized series reversion back to the fractional as described by Eq. \eqref{eq:ARFIMA}. The estimated scaling parameters for the reverted series are almost the same as for the original empiric, but with the lower standard deviation; see table \ref{table2} and Fig. \ref{fig5}. The finding serves as one more argument that order disbalance time series are self-similar as FLSM and ARFIMA processes suggest. Previously defined persistence \cite{Lillo2004SNDE,Bouchaud2004QF,Toth2015JEDC,Gontis2020JStatMech} is probably related to the power-low PDFs of order sizes than to the long-range memory effects.

It is important to know how various estimators of self-similarity and memory effects work, with FLSM and ARFIMA models being the most general theoretical concept of self-similar processes. The Burst duration analysis considered in \cite{Ding1995PhysRevE,Gontis2017PhysA,Gontis2017Entropy,Gontis2018PhysA,Gontis2020JStatMech} is important here as the burst duration has straightforward relation to the properties of first passage time used in many practical applications \cite{Metzler2014Springer,Metzler2019JStatMech,Padash2019JPhysA,Grebenkov2020JPhysA}. 
The shift of $H$ estimated from BDA in empirical and randomized order disbalance series gives the higher value of correlation parameter $d=-0.2$ than AVE and Higuchi's methods. Nevertheless, this method and MSD confirm the presence of anti-correlation in the original empirical order disbalance series. 
 Andersen's theorem has been confirmed by extensive numerical simulations of the first-passage time PDF \cite{Koren2007PhysA} of symmetric L\'evy flights within a Langevin dynamic approach. Nevertheless, we must admit that the BDA of the considered randomized empirical time series gives us considerably higher values of $H$ than expected $0.5$. More detailed theoretical and numerical investigation of burst duration and first passage time statistics for the FLSM and discrete ARFIMA processes with non-Gaussian noise is needed. A theoretical investigation should help explain the observed contradiction with Andersen's theorem in the case of uncorrelated increments in empirical series. 

\section*{References}
\bibliography{elsarticle-ARFIMA-Gontis}
\end{document}